\pdfoutput=1
\documentclass[aps,prd,amsmath,floats,floatfix, twocolumn,
superscriptaddress,nofootinbib,showpacs]{revtex4-1}

\usepackage[T1]{fontenc}
\usepackage[utf8]{inputenc}
\usepackage{lmodern}
\usepackage{verbatim}

\usepackage[dvipsnames, usenames]{xcolor}
\definecolor{linkcolor}{rgb}{0.0,0.3,0.5}
\usepackage[hypertexnames=false, unicode, colorlinks=true, linkcolor=linkcolor,
citecolor=linkcolor, filecolor=linkcolor,urlcolor=linkcolor,
pdfusetitle]{hyperref}
\usepackage{orcidlink}

\usepackage[all]{hypcap}
\usepackage{graphicx}
\usepackage{xspace}
\usepackage{amssymb}
\usepackage[normalem]{ulem} 
\usepackage{bm} 
\usepackage{enumitem,amssymb}

\usepackage{microtype}

\usepackage[english]{babel}
\usepackage{blindtext}

\graphicspath{%
  {figs/}%
}

\DeclareMathAlphabet{\mathpzc}{OT1}{pzc}{m}{it}

\newlist{todolist}{itemize}{2}
\setlist[todolist]{label=$\square$}

\newcommand{\rhodm}{\rho_{\rm DM}}
\newcommand{\gbl}{g_{B\!-\!L}}
\newcommand{\qbl}{q_{B\!-\!L}}
\newcommand{\bl}{B\!-\!L}
\newcommand{\dbl}{\Delta_{B\!-\!L}}

\newcommand{\Tobs}{T_{\rm obs}}
\newcommand{\Tcoh}{T_{\rm coh}}
\newcommand{\Teff}{T_{\rm eff}}

\newcommand*{\change}[1]{{\color{black}{#1}}}
\newcommand*{\new}[1]{{\color{black}{#1}}}
\newcommand*{\newer}[1]{{\color{black}{#1}}}

\begin{document}

\title{Differential torsion sensor for direct detection of ultralight vector dark matter}

\author{Ling~Sun\,\orcidlink{0000-0001-7959-892X}}
\affiliation{OzGrav-ANU, Centre for Gravitational Astrophysics, College of Science, The Australian National University, Australian Capital Territory 2601, Australia}
\email{ling.sun@anu.edu.au}

\author{Bram~J.~J.~Slagmolen\,\orcidlink{0000-0002-2471-3828}}
\affiliation{OzGrav-ANU, Centre for Gravitational Astrophysics, College of Science, The Australian National University, Australian Capital Territory 2601, Australia}
\email{bram.slagmolen@anu.edu.au}

\author{Jiayi~Qin\orcidlink{0000-0002-7120-9026}}
\affiliation{OzGrav-ANU, Centre for Gravitational Astrophysics, College of Science, The Australian National University, Australian Capital Territory 2601, Australia}
\email{jiayi.qin@anu.edu.au}


\hypersetup{pdfauthor={Sun et al.}}

\date{\today}

\begin{abstract}
Ultralight bosons with masses in the range from $\sim 10^{-22}$~eV/$c^2$ to $\sim 1$~eV/$c^2$, are well-motivated, wave-like dark matter candidates. Particles on the lower-mass end are less explored in experiments due to their vanishingly small mass and weak coupling to the Standard Model. 
We propose a sensor with dual torsion pendulums for the direct detection of U(1)$_{\bl}$ gauge boson dark matter, which can achieve an enhanced differential torque sensitivity in a frequency band of $\sim 10^{-2}$--10~Hz \change{due to its advantages in common-mode rejection and differential angular sensitivity}.
We describe the design of the differential torsion sensor and present the estimated sensitivity to an ultralight dark matter field coupled to baryon minus lepton ($\bl$) number, in a mass range of $\sim 10^{-17}$--$10^{-13}$~eV/$c^2$. 
Given a setup with meter-scale torsion pendulum beams and kg-scale test masses, the projected constraints on the coupling constant \new{$\gbl$} can reach \new{$\sim 10^{-27}$} for a boson mass of $\sim 10^{-15}$~eV/$c^2$. 
\end{abstract}
\maketitle

\section{Introduction}

Dark matter is an essential component in astrophysics and modern cosmology~\cite{Planck2018,Bertone2018}, and strong evidence for its existence has been observed from the gravitational behavior in the Universe, e.g., through gravitational lensing effects~\cite{Clowe_2006,Massey2010}. Nevertheless, the nature of dark matter remains elusive. The masses of dark matter candidates span over 90 orders of magnitude~\cite{Bertone2018a}, ranging from fuzzy dark matter (wave-like ultralight particles) with masses $\sim 10^{-22}$~eV/$c^2$~\cite{Hu2000,Hui2017}, weakly interacting massive particles (WIMP) with masses in the $\sim$~GeV/$c^2$--TeV/$c^2$ regime~\cite{Goodman1985}, to massive primordial black holes~\cite{Carr2010}. 
The wide spread in the mass range sets a challenge to detect the dark matter signature and requires various types of exquisitely sensitive experiments, probing different interaction mechanisms. 

The lower-end mass regime is particularly difficult to explore due to the particle's vanishing mass and is less tested with experiments.
In recent years, searches for ultralight dark matter have been proposed or carried out with experiments on various scales, \change{ranging from atomic clocks~\cite{Arvanitaki2015,Kennedy2020}, optomechanical cavities and laser interferometers~\cite{DeRocco2018,Obata2018,Liu2019,Martynov2020,Geraci2019, Pierce2018,Carney_2021,Manley2021}, including kilometer-scale ground-based gravitational-wave detectors~\cite{Vermeulen2021,Michimura2020,Guo2019,Nagano2019,Nagano2021,Abbott2022,O3GK,Morisaki2021} and LISA Pathfinder~\cite{Miller2023}, torsion-balance accelerometers~\cite{Schlamminger2008,Wagner_2012,Graham2016,Shaw2022}, and astrophysical approaches with black hole superradiance and pulsar timing~\cite{Sun2020,Abbott-O3-all-sky,Khmelnitsky_2014,Porayko2014}.}

Torsion balance, or torsion pendulum, is a key instrument for measuring extremely weak forces and has been widely used and improved over decades for tests of the equivalence principle~\cite{Adelberger1990,Smith1999,Schlamminger2008,Wagner_2012,Graham2016,Eot-wash}.
Ultralight vector dark matter field coupled to the Standard Model via the baryon minus lepton ($\bl$) number (i.e., neutron numbers) exerts time-oscillating, equivalence-principle-violating forces on normal matter. 
A recent study~\cite{Shaw2022} used a single torsion balance design equipped with four beryllium (Be) and four aluminium (Al) test masses to measure the torque from a $\bl$ coupled vector dark matter field. The search put a sensitive constraint on the coupling constant $\gbl (\hbar c)^{-1/2} < 1\times 10^{-25}$ for a boson mass of $\sim 8\times 10^{-18}$~eV/$c^2$, where $\hbar$ is the reduced Planck constant, and $c$ is the speed of light.\footnote{\new{In this paper, we use the convention to write $\gbl$ in units of $(2e^2/\varepsilon_0)^{1/2}$, where $e$ is the elementary charge and $\varepsilon_0$ is the vacuum permittivity. Note that Ref.~\cite{Shaw2022} reports $\gbl$ in units of $(\hbar c)^{1/2}$ instead.}}
\change{Future torsion balance experimental setups are proposed with an estimated reach of $\gbl < 1\times 10^{-31}$ in the mass range of $\sim  10^{-23}$--$10^{-16}$~eV/$c^2$ for a $\bl$ coupled vector field~\cite{Graham2016}. However, such aggressive future setups require various technology upgrades, e.g., using polypropylene test masses and silica fibers and operating in a cryogenic environment~\cite{Graham2016}.}

In this paper, we propose a novel optomechanical differential torsion sensor, an alternative design to the single torsion balance, \change{which takes advantage of a large common-mode rejection and its differential angular sensitivity.
Without requiring significant upgrades in technology,} the sensor can probe vector dark matter with a $\bl$ coupling constant \new{three} orders of magnitude lower than the \change{regime excluded in existing experiments}.
The core of the device, based on two torsion oscillators, was originally proposed for the detection of gravitational waves by Braginsky in 1969~\cite{misner2017gravitation}. Such sensors have been implemented, with test masses made of the same material, to measure low-frequency gravitational waves~\cite{Ando2010} or to measure gravitational field fluctuations to mitigate Newtonian noise in terrestrial laser interferometric gravitational-wave detectors~\cite{McManus_2017,Chua2023}.
Here, we describe \change{a new dark matter detection experiment, using the mid-scale differential torsion sensor}, equipped with two meter-size pendulum beams and kg-size equal-mass test masses made from different materials (e.g., Be and Al) on the two ends of each beam. 
We show that differential torsion sensors, advantageous in common-mode rejection and differential angular sensitivity, are sensitive to ultralight vector dark matter, capable of probing a coupling constant at the level of \new{$\gbl \sim 10^{-27}$} in the most sensitive frequency band of $\sim 10^{-2}$--10~Hz, corresponding to a mass range of $\sim 10^{-17}$--$10^{-14}$~eV/$c^2$.


\section{Vector dark matter field}

The Lagrangian of a massive vector dark matter field coupled to a number current density $J^\mu$ of baryons ($B$) or baryons minus leptons ($\bl$) is given by~\cite{Michimura2020}
\begin{equation}
    \mathcal{L} = -\frac{1}{4}F^{\mu\nu}F_{\mu\nu} + \frac{1}{2} m_A^2 A^{\mu}A_{\mu} - \epsilon e J^{\mu} A_{\mu},
\end{equation}
where $m_A$ is the mass of the vector field, $A_\mu$ is the four-vector potential of the field, $F_{\mu\nu} = \partial_\mu A_\nu - \partial_\nu A_\mu$ is the electromagnetic field tensor, $e$ is the elementary charge, and $\epsilon$ is the gauge coupling constant normalized by the electromagnetic coupling constant.

The dark matter orbiting the galactic center is non-relativistic, \change{with a virial velocity of approximately $220$~km/s around the solar system~\cite{Evans2019}.
The typical velocity of the dark matter, $\bar{v}$, at the surface of the Earth is dominated by a random virial velocity following the probability distribution of the standard halo and the velocity of the Sun with respect to the rest frame of the halo; the effect from the Earth's motion is negligible.}
The total energy of a dark matter particle with mass $m_A$ is the sum of its rest energy and kinetic energy, i.e., \change{$m_A c^2 [1 + (\bar{v}/c)^2/2]$, with $\bar{v}/c$ at the level of $\sim 10^{-3}$~\cite{Abbott2022,Smith2007,Nakatsuka2023}.}
The oscillation frequency of the field is approximately 
\begin{equation}
    f_0 = \frac{m_A c^2}{2 \pi \hbar}.
\end{equation}
There is a small fractional deviation on the order of $10^{-7}$, i.e., 
\begin{equation}
    \frac{\Delta f}{f_0} = \frac{1}{2}\left(\frac{\change{\bar{v}}}{c}\right)^2. 
\end{equation}
Thus, the local dark matter field can approximately be treated as a plane wave, 
\begin{equation}
    A_\mu \approx A_{\mu,0} \cos(2\pi f_0 t - \bm{k} \cdot \bm{x} + \phi), 
\end{equation}
where $A_{\mu,0}$ is the amplitude of the field, $\bm{k}$ is the propagation vector, $\bm{x}$ is the position vector, and $\phi$ is a random phase. 
In the non-relativistic regime, the time component $A_t$ is negligible compared to the spatial component $\bm{A}$, and hence we ignore $A_t$.
Here, the kinetic energy contribution is neglected, and the polarization and propagation are treated as constant vectors. Such an approximation is only valid within a duration \change{on the order of $1/\Delta f$} when the field is considered coherent.
We can write the dark matter electric field as
\begin{eqnarray}
    \nonumber \bm{E}_A &=& \partial_t \bm{A}(t, \bm{x}), \\ 
     &=& E_0 \hat{\bm{e}}_A \sin(2\pi f_0 t - \bm{k} \cdot \bm{x} + \phi),
\end{eqnarray}
where $\bm{A}(t, \bm{x})$ is the spatial vector field at time $t$ and position $\bm{x}$, $E_0$ is the amplitude of the dark matter electric field, and $\hat{\bm{e}}_A$ is the unit vector parallel to $\bm{E}_A$.
In the non-relativistic regime, the dark matter magnetic field is negligible compared to the dark matter electric field, and $E_0$ can be determined by 
\begin{equation}
    E_0 = \sqrt{2 \rhodm},
\end{equation}
where \change{$\rhodm \approx 0.4$~GeV/cm$^3$} is the local dark matter energy density~\cite{Tanabashi2018}.

In this paper, we consider a U(1)$_{\bl}$ gauge boson.
The field interacts with objects carrying $\bl$ ``charges'', $\qbl$, with a coupling constant $\gbl$.
For an object of a particular atomic species with a total mass $M$, the $\bl$ charge is
\begin{equation}
    \qbl = \frac{A-Z}{\mu}\frac{M}{m_n},
\end{equation}
where $A$ is the atomic mass number, $Z$ is the atomic number, $\mu$ is the atomic mass in atomic mass units, and $m_n$ is the neutron mass. The term ${(A-Z)}/{\mu}$ stands for the neutron-to-mass ratio (i.e., charge-to-mass ratio). 
Analogous to the electric force, the dark matter electric force on a free-falling object with a charge $q_{B-L}$ can be written as 
\begin{equation}
    \bm{F}_A = \gbl \qbl \bm{E}_A.
\end{equation}



\begin{figure}[!tbh]
	\centering
	\includegraphics[width=\columnwidth]{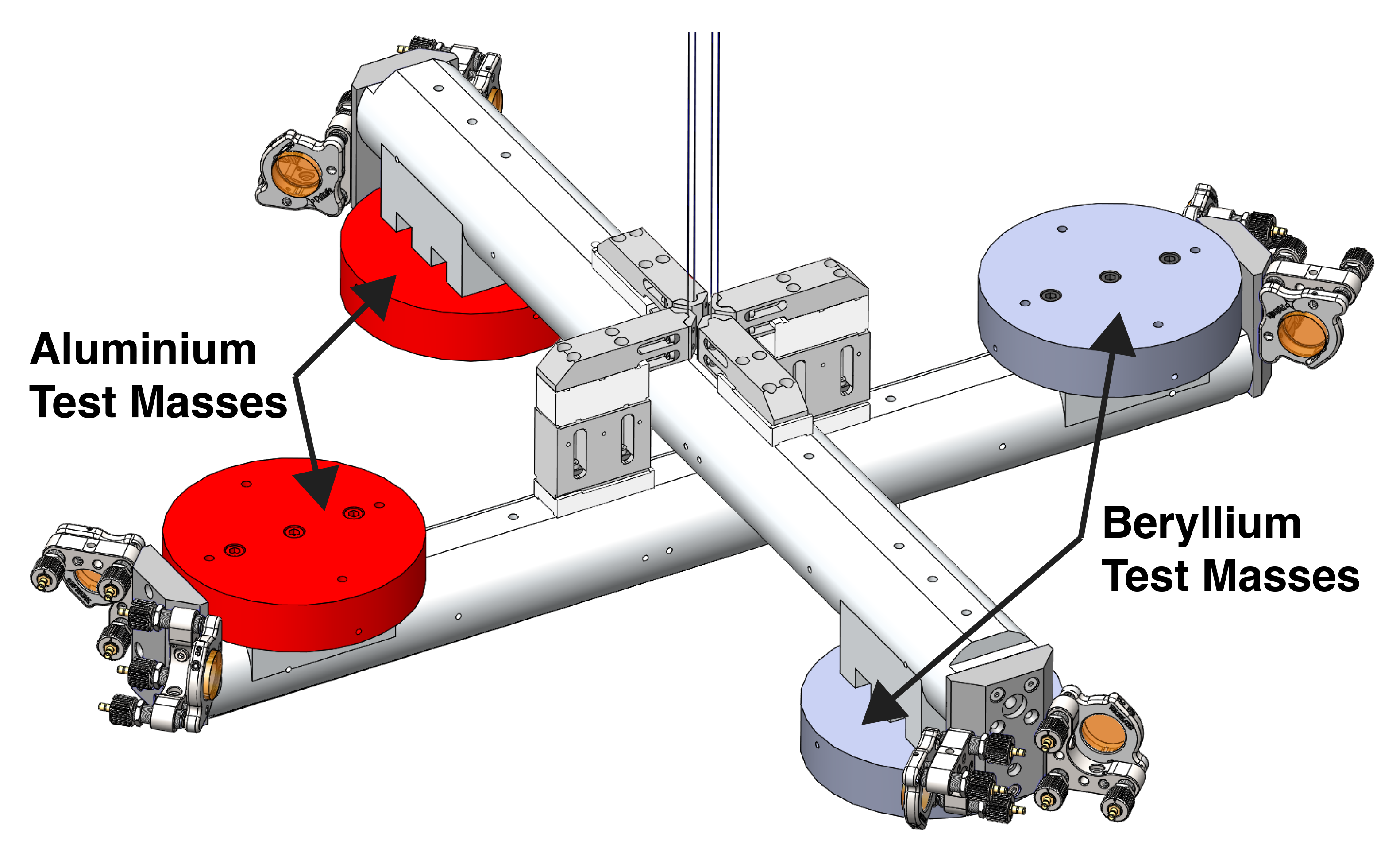}
	\caption[]{Conceptual design of the differential torsion sensor. The two pendulum beams are in a cross formation. The beryllium (Be) and aluminium (Al) test masses are highlighted at the end of each beam. \change{To sense the differential rotation,} four optical cavities are formed between the dual torsion pendulums with partial-reflecting mirrors (shown in orange colour). The shape and size of the test masses and beams are for demonstration purposes.}
	\label{fig:torpedo_bars}
\end{figure}


\section{Differential torsion sensor}

The differential torsion sensor consists of two independent torsion pendulum beams, suspended in a cross formation (see Figure~\ref{fig:torpedo_bars}). The centers of mass of the two pendulums are designed to overlap in the three-dimensional space.  
Each dumbbell-like torsion pendulum consists of a meter-size beam with kg-size test masses on both ends.
Each beam is suspended by two tungsten suspension wires that are attached to each pendulum above their centers of mass (at the top of the center grey blocks in Figure~\ref{fig:torpedo_bars}). 

The vector field is weakly coupled to the Standard Model via the $\bl$ number. 
The force strength that a free-falling massive object experiences depends on the number of neutrons (i.e., of $\qbl$ charges) within the object, which are different per unit mass of the two atomic species, Be and Al. 
The difference in the charge-to-mass ratios of Be and Al is
\begin{equation}
    \dbl = \frac{A_{\rm Be}-Z_{\rm Be}}{\mu_{\rm Be}} - \frac{A_{\rm Al}-Z_{\rm Al}}{\mu_{\rm Al}} = 0.0359,
\end{equation}
where the subscripts denote the atomic species. 
The dark matter field exerts a different force on the Be test masses compared to the Al test masses, resulting in a net torque. 
\change{In principle, other atomic species can also be used, e.g., tungsten (W) has a larger $\bl$ number than Be and can lead to better sensitivity. However, the larger density of W compared to Be would require more sophisticated mechanical engineering to make test masses with the same external dimensions. For practical reasons, we choose Be and Al test masses, as used in most of the other existing torsion balance experiments~\cite{Graham2016,Shaw2022}.}
The differential torsion sensor is sensitive to the vertical component of the differential torque, 
\begin{equation}
    \delta \tau_A = \bm{F}_A \cdot (\hat{\bm{z}} \times \bm{r}) = \frac{M}{m_n} \dbl \gbl \bm{E}_A \cdot (\hat{\bm{z}} \times \bm{r}),
    \label{eq:torque_dm}
\end{equation}
where $\bm{r}$ is the lever arm vector.

As shown in Figure~\ref{fig:torpedo_bars}, optical Fabry-P\'{e}rot cavities are used for measuring the differential rotation between the two pendulums, with partial-reflecting mirrors placed at the ends of each beam. A differential rotation of the pendulums causes common expansion and contraction of geometrically opposite cavities. Measuring an appropriate combination of cavity length changes gives the differential rotation readout of the sensor.
The Pound-Drever-Hall locking method can be used to obtain the cavity length error signal~\cite{Drever1983,Black2001}. The error signal is fed back to the laser to maintain cavity resonance and is used to reconstruct the differential rotation. 
\change{With each pendulum having six degrees of freedom, the sensing and control matrices are measured to minimize the sensing and actuation coupling of the other modes into the differential rotation.}

\begin{figure}[!tbh]
	\centering
	\includegraphics[width=\columnwidth]{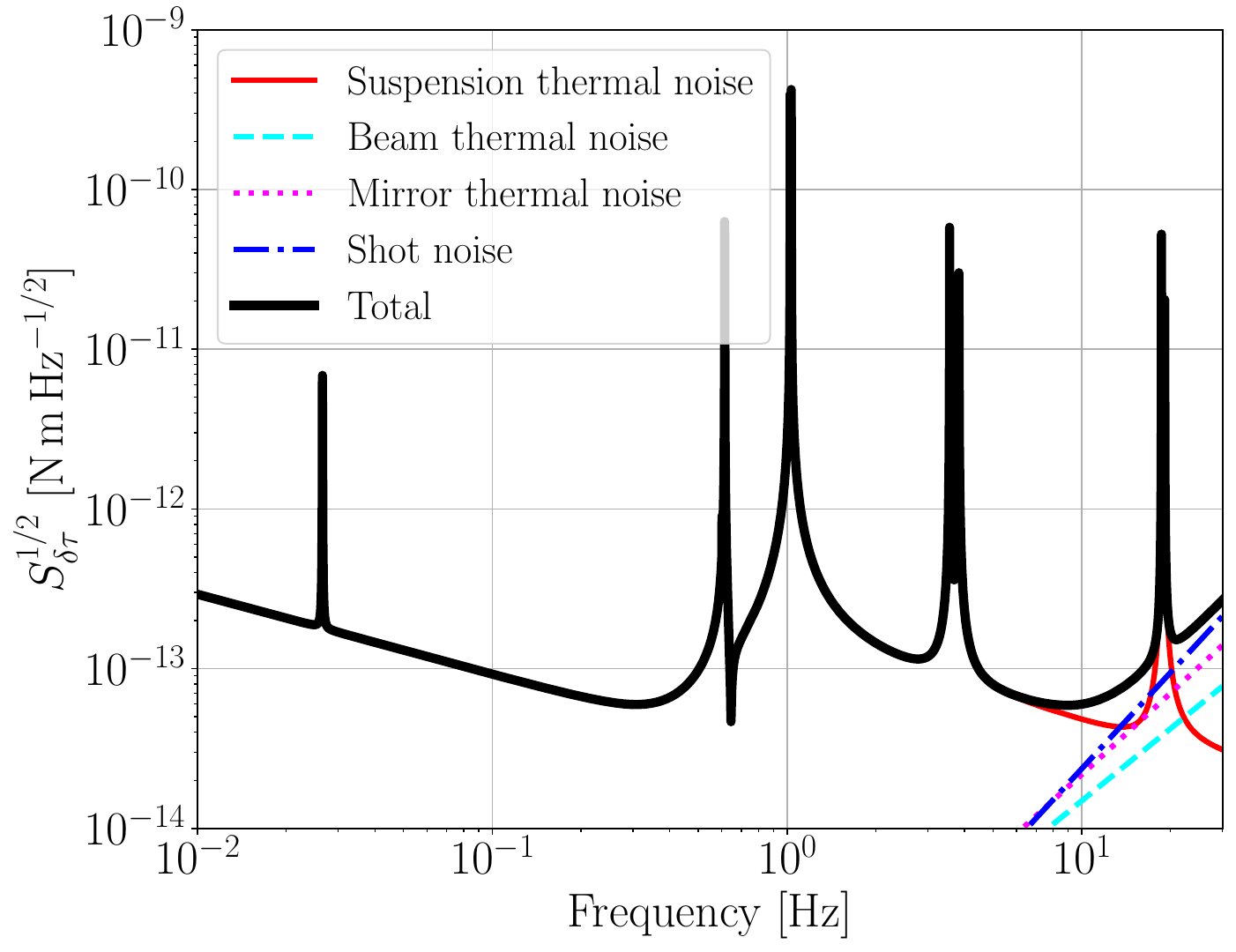}
	\caption[]{Projected amplitude spectrum density of the sensor's differential torque sensitivity \change{(thick black curve). The sensitivity is limited by suspension thermal noise (red solid curve) at lower frequencies~\cite{Chua2023} and by the combination of beam thermal noise (cyan dashed curve), mirror thermal noise (magenta dotted curve), and shot noise (blue dash-dot curve) at higher frequencies. The five major groups of spikes, from left to right, are caused by the effective difference between the two torsion resonances at around 0.026~Hz, longitudinal and transverse resonances, roll resonances, pitch resonances, and vertical resonances, respectively.} (See text for the fiducial configuration.)
 }
	\label{fig:diff_torque_spec}
\end{figure}

Assuming 5-kg test masses on both ends of each \change{0.6-m} sensor beam, we have a total moment of inertia of 0.75~kg\,m$^2$ for each pendulum, equivalent to having a $\sim 5$ kg sensor beam with additional mounting equipment and a mechanical lever arm from the axis of rotation to the center of each test mass of 0.24~m.
The torsion pendulums (suspended by two 0.6-m long tungsten suspension wires with a \new{$Q$ factor of $3\times 10^3$~\cite{Cavalleri2009.CQG}}) have a mechanical resonant frequency of $0.026$~Hz. 
\newer{The design incorporates a common suspension point of the two pendulums and provides a large suppression on mechanical common modes. The four-stage mechanical suspension and isolation chain ($1.5$~m in diameter and $3$~m tall), including low-noise sensors (such as Nanometrics Trillium 240~\cite{Trillium240}) and active feedback and controls, operating in a vacuum tank, can achieve a suspension point motion of} $10^{-11}$~m\,Hz$^{-1/2}$ and $10^{-11}$~rad\,Hz$^{-1/2}$ at $0.1$~Hz~\cite{Chua2023}.
This isolation performance guarantees a suspension thermal noise limited differential angular sensitivity of $6.7\times10^{-13}$~rad\,Hz$^{-1/2}$ at 0.1~Hz at a temperature of 300~K.
We consider an impedance-matched cavity with finesse of $\sim150$.
With an incident laser power of 10~mW at a wavelength of $1.064\,\mu{\rm m}$ and using the optimal modulation depth, the displacement-equivalent shot noise in the cavities is $4 \times 10^{-18}$~m\,Hz$^{-1/2}$. 
With a 0.24-m mechanical lever arm, the shot noise limited differential angular sensitivity is at the level of $10^{-17}$~rad\,Hz$^{-1/2}$.
\newer{Key formulae for suspension thermal noise and shot noise are provided in Appendix~\ref{appendix:formulae}.}
Both the beam vibration thermal noise and the mirror thermal noise are subdominant. 
Using beam tubes made of Niobium (with the first vibration mode frequency at $\sim 480$~Hz and a $Q$ factor of $10^{6}$~\cite{duffy_acoustic_1994}) and with a laser spot size of 1-mm diameter on the fused silica cavity mirrors, both the beam vibration thermal noise and the mirror substrate and coating thermal noise levels~\cite{Braginsky2003.PLA, Levin1998} are below the suspension thermal noise at lower frequencies and shot noise at higher frequencies.
With this setup, the projected amplitude spectrum density of the differential torque at design sensitivity, denoted by $S_{\delta \tau}^{1/2}$, is presented in Figure~\ref{fig:diff_torque_spec}~\cite{Chua2023}. 

\section{Sensitivity}
Next, we estimate the sensitivity to the $\bl$ coupled dark matter field using a fiducial configuration in the sensor design and derive the projected constraints on the coupling constant $\gbl$.

Following Equation~\eqref{eq:torque_dm}, the time-varying differential torque on average is 
\begin{equation}
   \sqrt{\langle \delta \tau_A^2 \rangle} \simeq \frac{M}{m_n} \dbl \gbl r \sqrt{\rhodm},
\end{equation}
where the root-mean-square amplitude of $\bm{E}_A$ (i.e., $E_0/\sqrt{2} = \sqrt{\rhodm}$) is used as the averaged field strength, and $r=0.24$~m is the lever arm length.
The signal-to-noise ratio (SNR) scales as
\begin{equation}
   {\rm SNR} = \frac{\sqrt{\langle \delta \tau_A^2 \rangle}}{\sqrt{S_{\delta\tau}(f)}} \sqrt{T_{\rm eff}},
\end{equation}
where $\Teff$ is the effective integration time. 
When the total observing time $\Tobs$ is shorter than the time over which the field can be treated as coherent, the signal power can be integrated fully coherently, and we have $T_{\rm eff} = \Tobs$. 
For long-duration observations, e.g., with $\Tobs \sim$ year, the total duration is usually divided into segments with a time length of $\Tcoh$, over which the signal power can be integrated coherently. The random phase only remains constant within $\Tcoh$, and the summation between different $\Tcoh$ segments is incoherent. 
In such a semicoherent integration, we have $T_{\rm eff} \simeq (\Tobs \Tcoh)^{1/2}$. 
Longer $\Tcoh$ increases the SNR given a fixed $\Tobs$.
By setting SNR to unity, we have a detectable limit on the differential torque, $S_{\delta\tau}^{1/2}(f) T_{\rm eff}^{-1/2}$. \change{In practice, the SNR required for a confident detection is usually at the level of 5--8, which depends on the noise statistics and false alarm probability. We do not discuss the SNR threshold here for brevity.}
Thus, we can derive the sensitivity on $\gbl$ from the differential torque sensitivity limit, via
\new{
\begin{equation}
    \gbl =\frac{m_n}{M \dbl r \sqrt{\rhodm}} \frac{\sqrt{S_{\delta\tau}(f)}} {\sqrt{T_{\rm eff}}}.
\end{equation}}

\begin{figure}[!tbh]
	\centering
	\includegraphics[width=\columnwidth]{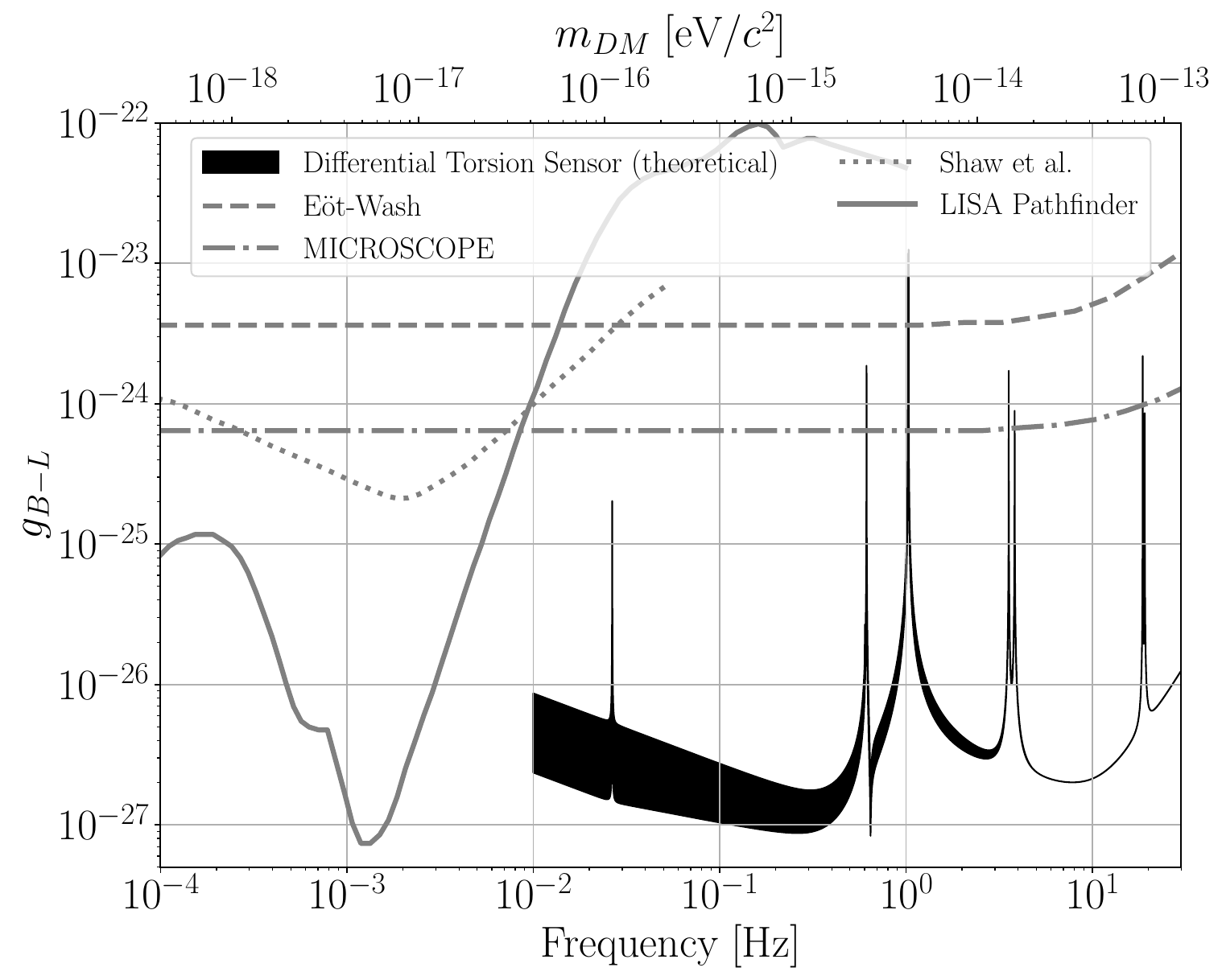}
	\caption[]{Theoretically projected searched sensitivity on the $\bl$ coupling constant (black band), assuming $\Tobs = 180$~day. The lower and upper bounds correspond to setting a coherent integration time using Equations~(\ref{eq:half_yr_coh}) and (\ref{eq:1day_coh}), respectively. 
    The existing limits set by the E{\"o}t-Wash experiment~\cite{Eot-wash}, MICROSCOPE~\cite{MICROSCOPE,Touboul2017}, the recent single torsion balance experiment by Shaw et al.~\cite{Shaw2022}, \newer{and LISA Pathfinder~\cite{Frerick2024}}, are plotted for comparison. }
	\label{fig:g_BL_sensitivity}
\end{figure}

We choose $\Tcoh$ by considering the maximum allowed coherent time at each frequency $f$,
\change{such that the signal power is concentrated in one frequency bin, $\Delta f_{\rm bin}(f)$, viz.,
\begin{equation}
    \label{eq:tau}
    \tau (f) = \frac{1}{\Delta f_{\rm bin}(f)} = \frac{1}{\kappa \left(\frac{\bar{v}}{c}\right)^2 f},
\end{equation}
where $\kappa \simeq 1.69$ is a factor that expands the bin size to cover the tail of the signal power distribution, and a typical value of $\bar{v} \simeq 1.2 \times 10^{-3}c$ is taken when considering the dark matter speed distribution~\cite{Nakatsuka2023}.}
\change{If we assume a total observing time of $\Tobs = 180$~day}, we have
\change{\begin{equation}
    \Tcoh(f) = {\rm min}\left[\tau(f), 180\, {\rm day} \right].
    \label{eq:half_yr_coh}
\end{equation}}
In practice, interruptions during observation due to, e.g., lock loss or maintenance, may prevent achieving the full estimated sensitivity over $\Tobs$. 
A more conservative choice of $\Tcoh$ is to limit the maximum coherent integration time within a day, given by
\change{\begin{equation}
    \Tcoh'(f) = {\rm min}\left[\tau(f), 1\,{\rm day} \right].
    \label{eq:1day_coh}
\end{equation}}
The theoretically projected search sensitivity for $\Tobs = 180$~day is shown in Figure~\ref{fig:g_BL_sensitivity} (black band). The lower and upper bounds correspond to setting $\Tcoh$ using Equations~(\ref{eq:half_yr_coh}) and (\ref{eq:1day_coh}), respectively.
Existing limits set by other experiments are shown for comparison. 

A more accurate sensitivity estimate is considered for future work, which should include as-built experimental implementations, software simulations, and rigorous analysis methods.
To achieve the desired sensitivity, other sources of noise apart from thermal noise and shot noise, e.g., Newtonian noise, magnetic noise, etc., need to be well monitored and controlled.
\newer{To ensure the magnetic noise is at least one order of magnitude below the noise budget, we require a field $\lesssim 100 \mu T$ (c.f. the environment background is at the level of $\sim 50 \mu T$).}
\newer{Magnetic noise can be further mitigated by placing the torsion beams in a cubic-meter-scaled enclosed shield made from mu-metal if necessary.} 
In addition, potential eddy currents are minimized by using low magnetic permeability materials \newer{close to} the torsion pendulums \newer{within the mu-metal enclosure}.
Newtonian noise, which cannot be shielded or attenuated, requires different techniques from mechanical isolation to be mitigated. 
\newer{We need one order of magnitude Newtonian noise mitigation with the experiment located in a low-seismic-noise environment equivalent to Peterson's new low-noise model ~\cite{Peterson1993.USGS}.}
There are multiple pathways to reduce the Newtonian noise to a subdominant level. An optimal solution is to build a second, closely positioned, identical differential torsion sensor equipped with the four test masses made of the same material, which will measure the Newtonian noise, allowing it to be canceled in the dark matter sensor. 
Other mitigation options include:
1) Placing the experiment underground can mitigate the impact of surface waves and body waves to some extent~\cite{Punturo2010}.
2) Excavating earth beneath the vacuum tank and replacing it with a lightweight fill can suppress local density and displacement fluctuations~\cite{Hall2021}.
3) A seismometer array can be effective in recording the seismic field in the vicinity of the differential torsion sensor and estimating the Newtonian noise contribution~\cite{Hall2021}.

In addition, uncertainties from both the potential signal and the instrument need to be accounted for. 
For instance, we essentially measure one sample of the stochastic amplitude of the field over a coherent duration, which may fall below the root-mean-square amplitude value. Other uncertainties involved in the experiment, e.g., measurement uncertainties and sensor calibration uncertainties, also need to be taken into consideration.
\change{In the absence of a detection, simulations and numerical calculations are required to derive robust constraints by accounting for the stochastic effects of the dark matter field and other statistical uncertainties, in a frequentist or a Bayesian approach.}

\section{Conclusion}
We propose using a novel differential torsion sensor for direction detection of $\bl$ coupled ultralight vector dark matter. 
Such an ultralight field, if it exists, would exert equivalence-principle-violating forces on normal matter, producing accelerations that the differential torsion sensor can measure. 
\change{Without relying on aggressive upgrades in technology required in future torsion balance experiments~\cite{Graham2016},} the proposed differential torsion sensor with a $\sim$~meter scale lever arm and $\sim$~kg size test masses \change{can achieve enhanced sensitivity to vector dark matter by taking advantage of the common-mode rejection and its differential angular sensitivity.}
We demonstrate that such a sensor is optimized for vector bosons in a mass range of $\sim 10^{-17}$--$10^{-14}$~eV/$c^2$, and the theoretically projected sensitivity can enable probing the coupling constant down to \new{$\sim 10^{-27}$, three orders} of magnitude lower than the \change{excluded regime} in published studies.
\new{Future upgrades, e.g., installing fused silica wires to enhance the Q factor, using heavier test masses, and extending beam lengths, will further increase the sensitivity of the experiment.}

\begin{acknowledgments}
The authors thank Perry Forsyth for providing transfer function data of the TorPeDO sensor prototype. 
This research is supported by the Australian Research Council Centre of Excellence for Gravitational Wave Discovery (OzGrav), Project Numbers CE170100004 and CE230100016. LS is also supported by the Australian Research Council Discovery Early Career Researcher Award, Project Number DE240100206.
\end{acknowledgments}

\appendix
\section{Useful formulae for noise estimates}
\label{appendix:formulae}

\newer{The power spectral density of suspension thermal noise (in units of m$^2$\,Hz$^{-1}$), as a function of angular frequency $\omega$, is derived from the fluctuation dissipation theorem~\cite{Callen1951,Callen1952} and is given by~\cite{Saulson1990,Cumming_2009,Cumming_2012}
\begin{equation}
S_{x}(\omega) = \frac{4 k_B T}{\omega^2} \Re(Y_x),
\end{equation}
where $k_B$ is Boltzmann's constant, $T$ is the absolute temperature, and $Y_x$ is the complex mechanical admittance of the torsion pendulum.}

\newer{The shot-noise-equivalent displacement of a cavity is given by~\cite{Black2001}
\begin{equation}
    S_{\rm shot} = \frac{1}{8 \mathcal{F}} \left( \frac{hc\lambda}{P_c}\right)^{1/2},
\end{equation}
where $\mathcal{F}$ is the finesse of the optical cavity, $h$ is Planck's constant, $c$ is the speed of light, $\lambda$ is the laser wavelength, and $P_c$ is the power in the carrier signal.}
\def\bibsection{\section*{References}}
%

\end{document}